\documentclass[]{aastex63}
\usepackage{hyperref}
\usepackage{wrapfig}
\usepackage{caption}
\usepackage{subcaption}
\usepackage{graphicx}
\usepackage{ragged2e}
\usepackage{wasysym}

\newcommand{\flu}{\mathcal{F}}

\newcommand{\dsn}{D}
\newcommand{\dkill}{D^{\rm kill}}

\def\iso#1#2{\mbox{${}^{#2}{\rm #1}$}}
\def\fe6#1{\iso{Fe}{6#1}}
\def\pu24#1{\iso{Pu}{24#1}}
\def\dop#1#2{\mathcal{W}^{#2}(#1)} 

\def\pfrac#1#2{\left( \frac{#1}{#2} \right)}

\begin{document}

\title{~~\\
Could a Kilonova Kill: a Threat Assessment\\
~~\\}

\author[0009-0000-5561-9116]{Haille M. L. Perkins}
\affiliation{Department of Astronomy, University of Illinois Urbana-Champaign, Urbana, IL 61801}
\affiliation{Illinois Center for Advanced Studies of the Universe, University of Illinois Urbana-Champaign, Urbana, IL 61801}
\affiliation{Center for Astrophysical Surveys, National Center for Supercomputing Applications, Urbana, IL 61801}
\author[0000-0002-7399-0813]{John Ellis} 
\affiliation{Theoretical Physics and Cosmology Group, Department of Physics, King's College London, London WC2R 2LS, UK}
\author[0000-0002-4188-7141]{Brian D. Fields}
\affiliation{Department of Astronomy, University of Illinois Urbana-Champaign, Urbana, IL 61801}
\affiliation{Illinois Center for Advanced Studies of the Universe, University of Illinois Urbana-Champaign, Urbana, IL 61801}
\affiliation{Department of Physics, University of Illinois Urbana-Champaign, Urbana, IL 61801}
\author[0000-0002-8028-0991]{Dieter H. Hartmann}
\affiliation{Clemson University, Department of Physics and Astronomy, Clemson, SC 29634, USA}
\author[0000-0002-8056-2526]{Zhenghai Liu}
\affiliation{Department of Physics, North Carolina State University, Raleigh, NC 27695, USA}
\author[0000-0001-6811-6657]{Gail C. McLaughlin}
\affiliation{Department of Physics, North Carolina State University, Raleigh, NC 27695, USA}
\author[0000-0002-4729-8823]{Rebecca Surman}
\affiliation{Department of Physics and Astronomy, University of Notre Dame, Notre Dame, IN 46556, USA}
\author[0000-0002-5901-9879]{Xilu Wang}
\affiliation{Key Laboratory of Particle Astrophysics, Institute of High Energy Physics, Chinese Academy of Sciences, Beijing, 100049, China}

\correspondingauthor{Haille Perkins}
\email{haillep2@illinois.edu}

\begin{abstract}
{Binary neutron star mergers (BNS)
produce high-energy emissions
from several physically
different sources, including a gamma-ray burst (GRB) and its afterglow, a kilonova, and, at late times, a remnant many parsecs in size. 
Ionizing radiation from these sources
can be dangerous for life on Earth-like planets when located too close. Work to date has explored the substantial danger posed by the GRB to on-axis observers: here we focus instead on the potential threats posed to nearby off-axis observers. Our analysis is based largely on observations of the GW~170817/GRB~170817A multi-messenger event, as well as theoretical predictions. For baseline kilonova parameters, we find that the X-ray emission from the afterglow may be lethal out to $\sim 5 \ \rm pc$ and the off-axis gamma-ray emission may threaten a range out to $\sim 4 \ \rm pc$, whereas
the greatest threat comes years after the explosion, from the cosmic rays accelerated by the kilonova blast, which can be lethal out to distances up to $\sim 11 \ \rm pc$. The distances quoted here are typical, but the values have significant uncertainties and depend on the viewing angle, ejected mass, and explosion energy in ways we quantify.
Assessing the overall threat to Earth-like planets,
have a similar kill distance to supernovae, but are far less common.
However, our results rely on the scant available kilonova data, and
multi-messenger observations will clarify the danger posed by such events.\\
  ~~\\
{KCL-PH-TH/2023-55, CERN-TH-2023-190}}
\end{abstract}

\keywords{Nucleosynthesis (1131), X-ray transient sources (1852), Gamma-ray transient sources (1853), Astrobiology (74), Supernovae (1668), Cosmic rays (329)} 

\section{Introduction} \label{sec:intro}

As the awareness and understanding of powerful cosmic transients have grown, so also has the realization of their dangers. The literature includes substantial discussions of the threat posed by nearby supernova (SN) explosions, gamma-ray bursts (GRB), and active galactic nuclei (AGN).  These events are copious emitters of ionizing radiation that may deplete stratospheric ozone, thereby subjecting the biosphere to large doses of ultraviolet radiation from the Sun \citep{mel_astrophysical_2011}.  At the worst, a sufficiently nearby outburst could trigger a biological mass extinction
\citep{Schindewolf1950,Schindewolf1954b,Krassovskij58,Shklovsky1969,Ellis:1993kc,Thomas2005, Fields2020}.

SN explosions emit X-rays and gamma radiation and subsequently accelerate lethal cosmic rays, and can be dangerous to life on Earth out to distances around $8-20 \ \rm pc$ \citep{Ellis:1993kc, Gehrels_2003, mel_astrophysical_2011, Thomas_2023}.  Moreover, if the circumstellar medium (CSM) surrounding a SN progenitor is sufficiently dense, the interaction between the explosion and the CSM can produce lethal X-rays that can kill out to $\sim 50 \ \rm pc$ \citep{brunton_x-ray_2022}. Being located on the axis of a jet from a GRB is potentially far more dangerous, as short GRBs (sGRBs), those with durations under $2 \ \rm s$, can provide a lethal dose of radiation up to $\sim 200 \ \rm pc$ away \citep{mel_astrophysical_2011}, increasing to $\sim 2 \ \rm kpc$ for long GRBs \citep{Thomas2005}. A study of the AGN phase of the Milky Way's Sgr A* found that planetary systems within $\sim 20 \ \rm pc$ of the central black hole could be left as barren rocks, as their atmospheres are stripped away by the ultraviolet and X-ray flux \citep{Chen2018}. Additionally, the accretion disk outflows from an AGN could render anywhere within $\sim 1 \ \rm kpc$ inhospitable for life \citep{Ambrifi2022} and the radiation could cause extinction in regions out to $\sim 3 \ \rm kpc$ from the largest active black holes \citep{lingam_active_2019}. 
These events play a role in shaping the expected Galactic Habitable Zone where conditions are thought to be favorable for Earth-like life \citep[e.g.,][]{Lineweaver_2004}. 

As the era of time-domain astronomy dawns, awareness is growing of new kinds of cosmic explosions, whose potential hazards merit study. This paper analyzes the consequences of nearby binary neutron star (BNS) mergers for their neighbors outside their jet beams and thus fortunate enough to be spared irradiation by the associated sGRBs.

The coalescence of a BNS or black hole-neutron-star (BHNS) system yields several distinct emission components. These produce emissions across the electromagnetic spectrum along with cosmic rays, and can last for timescales of a few seconds to years to millennia, as discussed in more detail in Section~\ref{sec:BNS-overview}. The first direct detection of a BNS merger was by the LIGO and Virgo collaborations via the gravitational waves produced from the event referred to as GW170817~\citep{LIGOScientific:2017vwq} and the accompanying electromagnetic emissions from GRB~170817A. These emissions were seen in many bands from radio to X-ray~\citep{LIGOScientific:2017zic}, and the GRB afterglow has been observed for over 4 years since the merger~\citep{OConner2022}.~\footnote{Such compact object mergers are predicted to be the main sources of {\em r}-process elements \citep{Lattimer1974, metzger_kilonovae_2020}, making these transients the targets of ongoing and upcoming observation campaigns that will produce many more measurements in the future.} 

It was suggested in~\cite{Ellis:1995qb} to search for live (not decayed) radioactive isotopes deposited on Earth by nearby supernova explosions, and recent geological and lunar data have made it clear that near-Earth explosions are an astrophysical reality \citep[for recent reviews see][]{FieldsWallner2023,Korschinek2023}. Live \fe60 signals have been measured by many groups in deep-ocean deposits \citep{Knie2004, Fitsoussi2008, Ludwig_2016, Wallner_2016, Wallner2021}, antarctic snow \citep{Koll_2019}, and in lunar regolith \citep{Fimiani_2016}.  These signals take the form of two pulses, one $\sim 3$ Myr ago and another $\sim 7-8$ Myr ago, pointing to at least two recent nearby SN explosions \citep{ertel_supernova_2023}.  Moreover, \citet{Wallner2021} detected live \pu244 in deep-ocean samples with high significance, confirming earlier hints \citep{Paul_2001, Wallner_2004, Raisbeck_2007, Wallner_2015}. As yet, the time resolution of the \pu244 signals is insufficient to tell if it they coincident with or independent from the \fe60 pulses. However, the presence of \pu244 is evidence for a recent nearby {\em r}-process event \citep{wang_r-process_2021, wang_proposed_2023, Wehmeyer2023}, which could be due either to a kilonova explosion prior to the formation of the local bubble, or production by one or more exceptional SNe within the Local Bubble. If the event was a kilonova, the measured \pu244 abundance suggests a distance of $\sim 300 \ \rm pc$ \citep{wang_r-process_2021}.  It is thus of particular interest to assess the threat posed by kilonovae at this range.

This paper is organized as follows: Section~\ref{sec:BNS-overview} describes the basic features of BNS mergers and their relevant emissions. Section~\ref{sec:ionizing-photons} highlights how various types of radiation endanger life on Earth. A brief review of the threat of sGRBs is given in Section \ref{sec:grb}. The threats imposed by BNS X-rays, gamma-rays, and cosmic rays are provided in Sections \ref{sec:xray}, \ref{sec:gammaray}, and \ref{sec:cosmicray}, respectively. The resulting overall threat of BNS mergers is assessed in Section~\ref{sec:rates}. Finally, in Section~\ref{sec:discussion}, we review the threat of binary neutron stars as found in this analysis and note the importance of future observations and theory.

\section{Overview of Binary Neutron Star Merger Features and Hazards} \label{sec:BNS-overview}

BNS mergers have several physically different components each producing unique and potentially threatening radiation. Figure~1 highlights each of these different components and their emissions. Fig.~1a is inspired by that of \citet{metzger_kilonovae_2020} and gives a zoom-in view around the time of the merger.  The jet orientation is away from the viewer, which is the more probable case, and that considered in this paper. 

The GW170817 gravitational wave event has helped to confirm that BNS mergers are the source of short GRBs \citep{abb_multi-messenger_2017}. When the GRB jet interacts with the interstellar medium, an afterglow of synchrotron emission is observable across the electromagnetic spectrum \citep{Makhathini_2021}. As we shall show, for the purposes of our analysis of the potential BNS threat from electromagnetic radiation, the X-ray emission is the most significant. In the case of GW170817, the afterglow has been observed for over 4 years \citep{OConner2022}.

The merger also yields dynamical ejecta including newly-synthesized {\em r}-process elements. These elements decay to produce the UV/optical/IR transient kilonova (KN) and were observed for $\sim$ 3/10/20 days after GW170817 \citep{abb_multi-messenger_2017}, respectively. While not yet observed, there has been a substantial amount of work done to predict the gamma-ray spectra of decaying {\em r}-process species. Many are predicted to produce MeV gamma rays through beta decay \citep{Korobkin_2020}, but certain actinide species may generate gamma rays with energies between $3-10 \ \rm MeV$ while fissioning \citep{wang_mev_2020}. 

After the ejecta material is thrown out, the sGRB attempts to pierce through it. When the jet interacts with the ejecta, it decelerates and creates a hot cocoon, seen in Figure~1a \citep{Ramirez-Ruiz2002, Nakar2017, Gottlieb_2017, Kasliwal2017}. This leads to two possible ways for gamma rays to be seen at larger observing angles than the jet, e.g., the cocoon can Thomson-scatter gamma rays from the jet \citep{Kisaka_2018}. In addition, \cite{gottlieb_cocoon_2018} found that as the shock in the cocoon breaks out of the ejecta, there is brief emission of high-energy photons. 

At late times, the ejecta from the merger blast will carve out a bubble in the interstellar medium that is thought to look much like that of the remnants generated by SNe \citep{Montes2016, Wu2019, Green2019}. This structure is shown in Fig.~1b. As with SN remnants, the merger remnant should accelerate cosmic rays.  As these engulf the surrounding stars, they will pose an additional hazard to Earth-like planets, potentially more threatening than the electromagnetic emissions from the BNS merger.

\begin{figure} \label{fig:zoominout}
    \centering
    \begin{subfigure}[t]{0.45\textwidth}
       \centering
       \includegraphics[width=\linewidth, page=1]{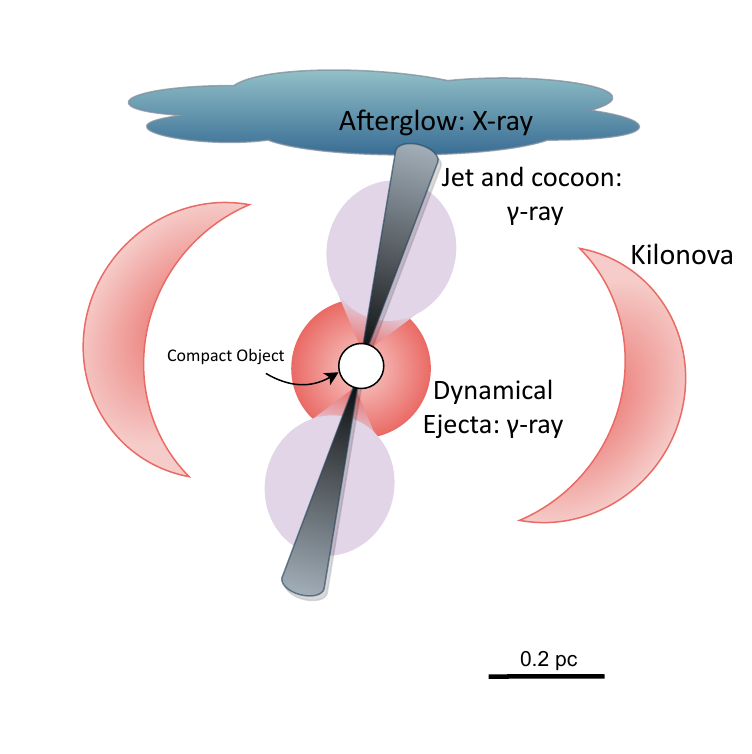}
       \vspace{-4em}
       \caption{First days to years post-merger}
   \end{subfigure}
   \hfill
   \begin{subfigure}[t]{0.45\textwidth}
       \centering
       \includegraphics[width=\linewidth, page=2]{cocoon.drawio.pdf}
       \vspace{-4em}
       \caption{Thousands of years post-merger}
   \end{subfigure}
    \caption{Schematic diagrams highlighting the interesting  components of emission from a binary neutron star merger at two different temporal and spatial scales. In (a), the system is shown at sub-parsec scales and within the first few years after the merger. The locations of gamma-ray emission in the dynamical ejecta and the cocoon surrounding the jet are highlighted, as well as the X-ray emission from the interaction between the jet and the interstellar medium (ISM). While the jet and cocoon emission are short-lived, the afterglow produces continuum emission for many years. Only the emitted X-rays produce the fluence that could be lethal. The kilonova, i.e., the UV/optical/IR emission powered by radioactivity in the ejecta, is also shown for additional context, but is undetectable after a week. 
     In (b), the larger, late-time structure of the remnant is shown with the long-since dissipated jet location indicated. The explosion from the merger will launch a strong shock that sweeps out a bubble-like structure as it expands through the ISM, which is a source of potentially threatening cosmic rays.}
\end{figure}

\section{The Threat from Ionizing Photons:  X-Rays and gamma rays} \label{sec:ionizing-photons}

Ionizing photons do not reach the Earth's surface, as they are stopped by the atmosphere. However, the atmospheric interactions can damage substantially the stratospheric ozone layer, thereby removing the biosphere's protection from the Sun's ultraviolet (UV) radiation. 
The ozone destruction is indirect:  $\rm O_3$ is generally not destroyed by the energetic photons themselves, but rather, ionizing particles interact with atmospheric nitrogen and oxygen to produce nitrates, mostly NO and $\rm NO_2$, collectively denoted ${\rm NO}_x$ \citep{brunton_x-ray_2022, Thomas2005}.  The nitrates destroy ozone in catalytic reactions so that one nitrate molecule can destroy more than one $\rm O_3$. 
\cite{Thomas2005} reviewed how the ozone loss could have lethal consequences to life, as the increased UVB would damage the DNA of living organisms.

\cite{ejzak_terrestrial_2007} found that only the total radiation dose that arrives within around $\sim 4 \ \rm years$ is significant, because the ozone layer will begin to recover after this period. On this timescale, the irradiation only depends on the integrated flux or {\em fluence}. For emission in band $b$ with flux $F_b(t)$, the fluence is
\begin{equation}
    \flu_b = \int F_b(t) \ dt \, .
\end{equation}
For an explosion at a distance $D$ with luminosity $L_b(t)$ in band $b$, the fluence is
\begin{equation}
\flu_b = \frac{\int L_b(t) \ dt}{4\pi D^2} = \frac{E_b}{4\pi D^2} \, ,   
\end{equation}
where $E_b = \int L_b(t) \ dt$ is the total energy emitted in band $b$.

It is important to consider separately different bands of radiation, because the energy possessed by the photon determines how deep it can penetrate into the atmosphere and whether it can reach the stratospheric ozone. Photons with energies $\lesssim 3 \ \rm keV$ are stopped high in the upper atmosphere and are unlikely to damage the ozone layer.  Photons with $3-10$ keV are mostly absorbed somewhat above the ozone layer, but atmospheric flows may carry some of the ${\rm NO}_x$ they produce down to ozone-bearing regions.  However, photon with energies exceeding $10 \ \rm keV$ are able to reach the ozone layer directly, so their nitrate production will be particularly effective in ozone destruction. 

When $\rm O_3$ damage is sufficient to cause a large reduction, this gives rise to a substantial increase in UVB radiation that in turn can lead to an extinction-level event. \cite{mel_astrophysical_2011} determined that $\geq 30\%$ global depletion of the ozone layer will have such consequences, and the fluence required to achieve this is known as the critical fluence, $\flu^{\text{crit}}_b$. The damage resulting from a given fluence level depends on the photon energy, and thus is different for each energy band $b$. \cite{brunton_x-ray_2022} lists three different levels of damage. 
The most significant damage comes from a fluence of $100 \ \rm kJ \ m^{-2}$ of gamma-ray photons ($E_{\gamma} \sim 200 \ \rm keV$), which \cite{mel_astrophysical_2011} quote as level required to achieve extinction-level ozone depletion. 
On the other hand a fluence of $200 \ \rm kJ \ m^{-2}$ of photons with energies near $10 \ \rm keV$ would achieve $30\%$ depletion, and a fluence of $400 \ \rm kJ \ m^{-2}$ of photons with a peak spectral energy near $2$ keV would inflict a similar level of damage~\citep{ejzak_terrestrial_2007}.

These estimates of the lethal fluences in different bands determine the largest distance from the event that is exposed to lethal radiation in each individual band $b$:
\begin{equation}
    \label{eqn:dleth}
    \dkill_b = \left( \frac{E_{b}}{4\pi w_b \flu^{\text{kill}}_b} \right)^{\frac{1}{2}} \, ,
\end{equation}
referred to as the `lethal distance'. In general there is emission across multiple bands, e.g., both X-ray and gamma-ray photons.
In this case, an event at distance $\dsn$ yields a total effective fluence on Earth of $\flu_{\rm eff} = \sum \flu_{{\rm eff},b}$, with $\flu_{{\rm obs},b}=w_b E_b/4\pi 
\dsn^2$ were $w_b \le 1$ measures the efficiency with which photons in band $b$ destroy stratospheric ozone.  The lethal fluence is defined such that $w_b \propto 1/\flu_b$, and the $w_0 = 1$ is maximum for gamma rays, which have the minimum lethal fluence $\flu_0 = 100 \ \rm kJ \ m^{-2}$.  Thus we can write
$w_b = \flu_0/\flu_b$, and we have an effective fluence at distance $\dsn$ of
\begin{equation}
    \flu_{\rm eff} = \sum w_b \flu_b = \frac{\sum w_b E_b}{4\pi \dsn^2} = \frac{\flu_0}{\dsn^2} \sum  \frac{E_b}{4\pi \flu_b}
    = \flu_0 \sum_b \pfrac{{\dkill_b}}{\dsn}^2 \, .
\end{equation}
With the bands weighted in this way, the lethality condition sets the observed effective fluence to be  $\flu_{\rm eff}= \flu_0$, which gives a net lethal distance
\begin{equation} \label{equ:dleth-tot}
    \dkill_{\rm net} = \sqrt{ \sum_b {\dkill_{\rm b}}^2} \, ,
\end{equation}
and we see that the net lethal distance is the sum in quadrature of the contributions from all bands.

\section{Dangers of Gamma Ray Bursts} \label{sec:grb}

As discussed above, BNS mergers are understood to be the main progenitors of short gamma ray bursts (sGRBs). These highly energetic events have been studied for their deleterious effects to terrestrial biospheres in \cite{mel_astrophysical_2011, Spinelli_2023}. A jet from a galactic GRB pointing towards Earth could disturb aquatic food webs leading to the death of a large portion of the phytoplankton that are an important food source in the oceans \citep{Guimarais_2010, Rodriguez_Lopez_2021}, and it has been suggested that a GRB may have been associated with the mass extinction event in the Ordovician period \citep{Melott_2004}. 
We provide a crude estimate of sGRB effects here in order to illustrate the transition to the off-axis kilonova case.

Since sGRBs exhibit relativistic beaming, the energy received and therefore the threat from these events depends sensitively on the viewing angle. We use the observed energy $E_{\rm obs}$
(i.e., time-integrated luminosity) of a relativistic source emitting isotropically with rest-frame energy $E'$ \citep{Rybicky_1982}:
\begin{equation}
    \label{eqn:Lobs}
    E_{obs} = \dop{\theta}{3}E' \, ,
\end{equation}
where $\mathcal{W}$ is an effective width with the same scaling as a Doppler factor:
\begin{equation} \label{eqn:dop}
    \dop{\theta}{} = \frac{1}{\Gamma (1 - \beta \cos \theta)} \, ,
\end{equation}
with $\beta = \frac{v}{c}$ and the Lorentz factor $\Gamma = (1-\beta^2)^{-1/2}$.  We use this as a crude means of demonstrating the strong effects of the viewing angle.

\begin{figure}
    \centering
    \includegraphics[width=\textwidth]{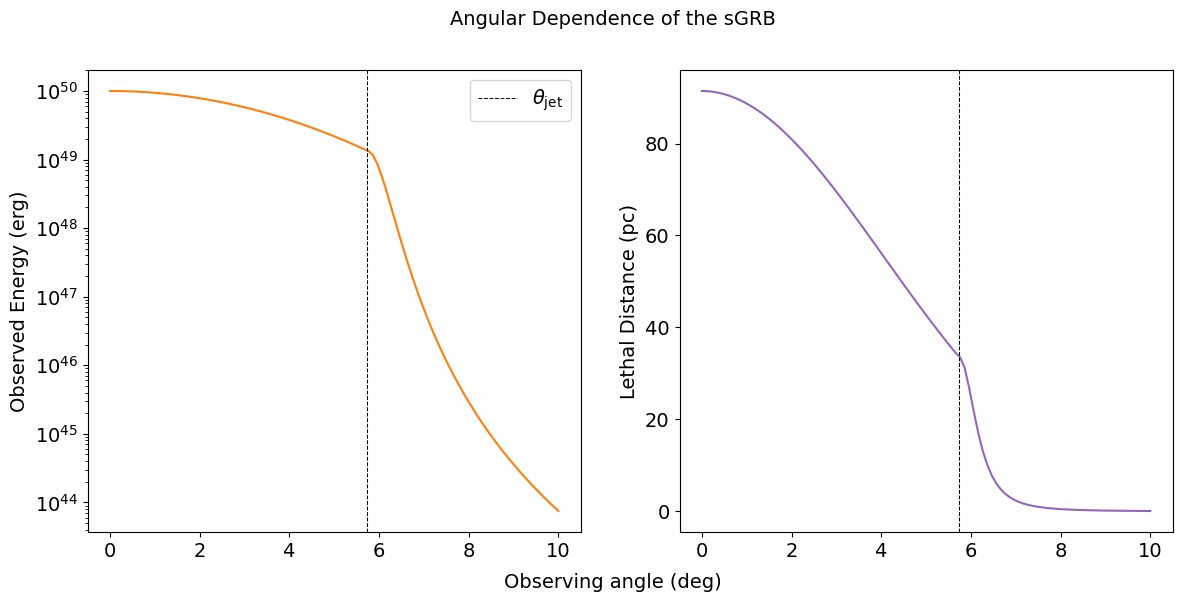}
    \caption{The left and right panels show, respectively, the sGRB energy and associated lethal distance as functions of the observing angle. These both assume $E_{\rm iso, 0} = 10^{50} \ \rm erg$, $\flu_{\gamma}^{\rm kill} = 100 \ \rm kJ \ m^{-2} $, and $\Gamma = 100$.}
    \label{fig:GRB-theta}
\end{figure}

GRB jets are likely to have a complex structure, with a flow profile that varies with the angle from the axis.  To incorporate a structured jet, we model the energy emitted by an sGRB with a piecewise function 
\begin{equation} \label{eqn:GRBmodel}
    E_{\rm GRB, iso}(\theta_{\rm v}) \propto \left\{ 
    \begin{array}{lll}
        e^{-\frac{1}{2}\left( \frac{\theta_v}{\theta_{\rm j}/2} \right)^2 } &,& \theta_{\rm v} \leq \theta_{\rm j} \nonumber \\
        \dop{\theta_{\rm v} - \theta_{\rm j}}{3} &,& \theta_{\rm v} > \theta_{\rm j} \nonumber \\
    \end{array} \right. ,
\end{equation}
where $\theta_{\rm v}$ is the viewing angle and $\theta_{\rm j}$ is the jet opening angle, which we take to be $0.1 \ \rm rad$ as is consistent with observations of GW170817 \citep[e.g.][]{Fong_2017, Mooley_2018, Beniamini_2019, metzger_kilonovae_2020}. We assume a Gaussian structure with on-axis observed isotropic energy $E_{\rm iso, 0}$ \citep{Hayes_2020} and width $\theta_{\rm j} / 2$. This model assumes that there is no beaming penalty until just outside of the jet. A calculation using Equation~(\ref{eqn:GRBmodel}) is shown in Figure~\ref{fig:GRB-theta}, using typical values for sGRBs, namely $\Gamma = 100$ and therefore $\beta = 0.99995$ \citep[e.g.][]{Ramirez-Ruiz2002} and $E_{\rm iso, 0} = 10^{50} \ \rm erg$ \citep{mel_astrophysical_2011}.
For an on-axis GRB, we get a lethal distance of $91 \ \rm pc$ when assuming a typical value for the lethal fluence of  gamma rays of $100 \ \rm kJ \ m^{-2}$. However, as found in \citep{ejzak_terrestrial_2007}, the ozone depletion can exceed out to a lethal distance $ \simeq 200 \ \rm pc$ the 30\% used to define the lethal gamma ray fluence if the energies of the photons are much greater than $200 \ \rm keV$  \citep{mel_astrophysical_2011}.

It is important to note that while we selected a typical sGRB energy, following \cite{mel_astrophysical_2011}, observations find a distribution of possible sGRB energies \citep{Salafia_2020}. Assuming constant luminosity and a duration of $2 \ \rm s$, the range of typical energies is on the order of $10^{46} - 10^{51}\ \rm erg$, corresponding to an on-axis lethal distance of $1 - 300 \ \rm pc$. However, as shown in Fig.~\ref{fig:GRB-theta}, the range of threat from the GRB drops immediately outside the jet opening angle. While the on-axis threat may vary, it is a general feature that the threat is negligible outside the jet, and the effects of the other BNS emission components become more significant than those of the sGRB, as described in the following Sections.

\section{The X-Ray Threat} \label{sec:xray}

\subsection{Afterglow of GW170817} \label{sec:afterglow}

Observations of GW170817 with NuSTAR and Swift \citep{evans_swift_2017} have found that the produced X-rays lie within the soft regime (below $10 \ \rm keV$). As indicated in \cite{brunton_x-ray_2022}, significant depletion of the ozone layer requires a greater deposition of energy from soft X-rays, and the lethal fluence for soft X-rays can be conservatively taken to be $400 \ \rm kJ \ m^{-2}$.

The jet interaction with the ISM produces an afterglow that radiates over longer timescales, and over much of the electromagnetic spectrum \citep[e.g.,][]{GRB_afterglow_review}; it is possible that interactions of the kilonova ejecta with the ISM could also produce their own afterglow emissions \citep{Kathirgamaraju_2019, Balasubramanian_2022}.  
\cite{Makhathini_2021} found that a broken power-law of the following form can be used to describe GW170817's afterglow across a broad range of frequencies:
\begin{equation} \label{eqn:Makhathini-model}
    F_{\nu}(t, \nu) = 2^{1/s} \bigg( \frac{\nu}{3 \text{GHz}} \bigg)^{\beta} F_p \bigg[ \bigg(\frac{t}{t_p}\bigg)^{-s\alpha_1} + \bigg(\frac{t}{t_p}\bigg)^{-s\alpha_2}\bigg]^{-1/s} \, .
\end{equation}
Using this model with the best-fit parameters listed in Table \ref{tab:params}, the total energy can be found by integration:
\begin{equation}
    E_{\rm X} = 4\pi \dsn^2 \int F_{\nu}(t, \nu) \, dt \, d\nu \, .
\end{equation}

\begin{table}[htb]
    \centering
    \caption{GW170817 Afterglow Flux Model Parameters from \cite{Makhathini_2021}}
    \label{tab:params}
    \begin{tabular}{c|cc}
Parameter & Value & Units \\
\hline
Peak Flux $F_p$ & $101 \pm 3$ & $\mu \rm Jy$ \\
Peak Time $t_p$ & $155 \pm 4$ & days \\
$\alpha_1$ & $0.86 \pm 0.04$ & \\
$\alpha_2$ & $-1.92 ^{+0.10}_{-0.12}$ & \\
$s$ & $3.6 ^{+1.0}_{-0.9}$ & \\
$\beta$ & $-0.584 \pm 0.002$ & \\
    \end{tabular}
\end{table}

For 900 days after the merger, emissions from GW170817 maintained a spectral index $\beta = -0.583 \pm 0.013$ \citep{hajela_evidence_2022, Fong_2019}, which is consistent with the parameterization of the spectrum in Equation~(\ref{eqn:Makhathini-model}). An observation that deviated from the afterglow model suggested that the kilonova afterglow, or interaction between the kilonova ejecta and the ISM, was beginning to emerge \citep{balasubramanian_continued_2021, hajela_evidence_2022, troja_accurate_2022}. However, more recent X-ray \citep{OConner2022} and radio \citep{Balasubramanian_2022} observations suggest that this is not the case, and that the jet afterglow is still visible out to 1671 days after the merger. These measurements also constrain the rebrightening from the kilonova afterglow to times beyond 4 years since the initial radiation, into the period of ozone recovery. In view of the expected timing and intensity of the rebrightening, we conclude that kilonova afterglow would not be a significant contributor to the lethal radiation.

The NuSTAR non-detections \citep{evans_swift_2017} between $3-10$ keV set upper bounds on the X-ray emission in this energy range. Knowing the index for the electron spectrum, the SWIFT observations can be extrapolated to higher energies. We have extrapolated the SWIFT observations closest in time to the non-detection by NuSTAR by assuming that the flux density $F_{\nu}$ at a frequency $\nu$ may be parametrized as
\begin{equation}
    \label{eqn:fluxdensity}
    F_{\nu} \propto \nu^{\beta} \, ,
\end{equation}
where the spectral index $\beta$ is found in \cite{Fong_2019}.
At $0.6$ and $4.5$ days post merger, measurements with SWIFT yielded flux densities of $0.0078$ and $0.0018 \ \rm \mu Jy$ at $1 \ \rm keV$ \citep{evans_swift_2017, Makhathini_2021}, respectively, whereas NuSTAR reported no detections at $0.7$ and $4.3$ days \citep{evans_swift_2017} within sensitivities for $6-10$ and $10-30$ keV of $2\times10^{-15}$ and $1\times10^{-14} \ \rm erg \ cm^{-2} \ s^{-1}$, respectively~\footnote{Sensitivity information can be found in version 3.2 of the NuSTAR Observatory Guide at \url{https://heasarc.gsfc.nasa.gov/docs/nustar/nustar_obsguide.pdf}.}. Integrating Equation~(\ref{eqn:fluxdensity}) over these energy ranges provides estimates of the fluxes in these energy bands, which are tabulated in Table~\ref{tab:swift-nustar}.

\begin{table}[htb]
    \centering
    \caption{Extrapolation of SWIFT observations into the NuSTAR observing range}
    \label{tab:swift-nustar}
    \begin{tabular}{c|c|c|c}
 Time (days) & $F_{\nu} (\mu \rm Jy)$ & $F_{6-10} \ (\rm 10^{-15} \ erg \ cm^{-2} \ s^{-1}) $ & $F_{10-30}  \ (\rm 10^{-14} \ erg \ cm^{-2} \ s^{-1})$ \\
\hline
0.6 & 0.0078 & 23 & 6.8 \\
4.5 & 0.0018 & 5.2 & 1.6 \\
    \end{tabular}
\end{table}

We find that the expected fluxes at these higher energies exceed the sensitivity limits in both of the bands listed by NuSTAR. Thus the non-detection by NuSTAR indicates that the maximum X-ray energy should be $\sim 6 \ \rm keV$. We note also that, as found in \citep{brunton_x-ray_2022}, photons with energies below $3 \ \rm keV$ are blocked by the upper atmosphere and therefore do not reach the ozone layer. 

The emission of interest is in the range $3-10 \ \rm keV$ before $4 \ \rm yr$ after the initial observation. Even though the $10 \ \rm keV$ upper limit may be an overestimate, it ultimately has little effect on our overall findings. The total energy in X-rays  and the corresponding lethal distance are 
\begin{eqnarray}
    E_{\rm X}(0.6 \ \rm rad) & = & 6.5 \times 10^{46} \ \rm erg \label{eqn:Ex}\\
    \dkill_{\rm X}(0.6 \ \rm rad) & = & 1.2 \ \rm pc \, .\label{eqn:Dxkill}
\end{eqnarray}

\subsection{The Effect of the Observing Angle} \label{sec:xray-angle}

To investigate how the angle of observation affects the observed X-ray afterglow energy, we assume it follows the same shape as the sGRB discussed in Section~\ref{sec:grb}. To account for the spreading of the jet as it interacts with the ISM, we use a wider jet opening angle than for a typical GRB. To do so, we assume that the area under the gaussian jet structure is conserved and that, in the non-spreading case, the jet has the typical values taken in Section~\ref{sec:grb} and the X-ray luminosity follows the canonical model described in~\citet{Nousek2006}. This model describes the temporal dependence of the luminosity by a piecewise function with 3 power-law segments. For a typical X-ray luminosity of $L_{\rm X,obs} \approx 8 \times 10^{43} \ \rm erg s^{-1}$ at $11 \ \rm hr$~\citep{De_Pasquale_2006, Berger_2014}, the segment of interest is well after the second break, which tends to occur at $\sim 1.5 \ \rm hr$. Using common values in the distribution of the temporal indices of the final segment, the on-axis luminosity is proportional to $t^{-1.3}$. Therefore, the on-axis energy $E_{\rm X, ns}(\theta = 0) \propto \int t^{-1.3} dt$. 

\begin{figure}
    \centering
    \includegraphics[width=\textwidth]{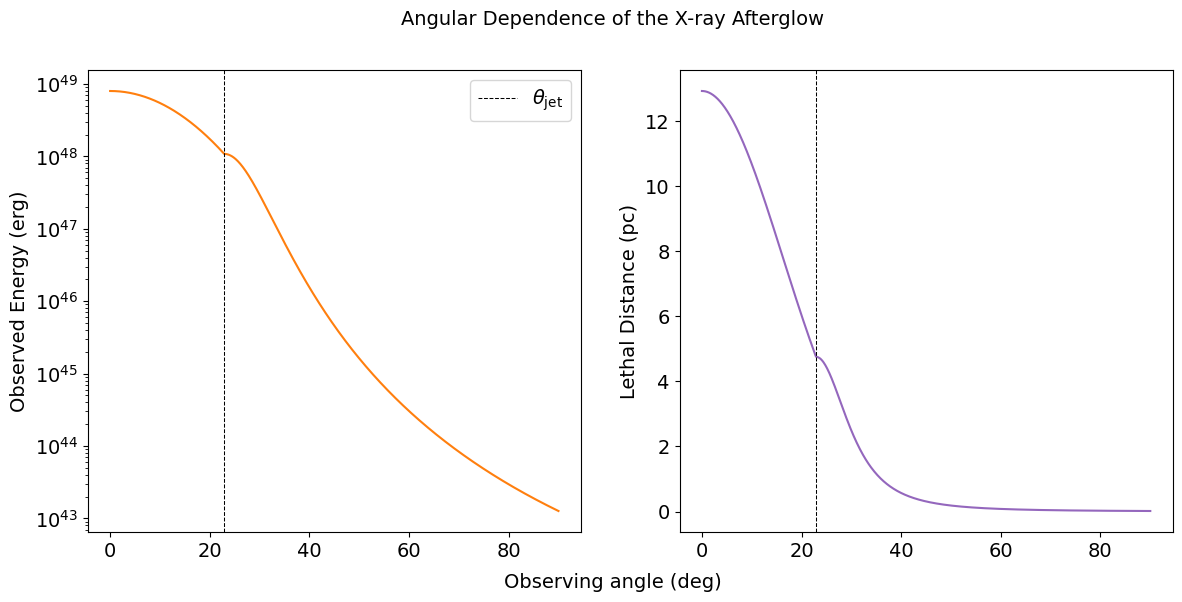}
    \caption{The left and right panels show the X-ray luminosity and associated lethal distance, respectively, as functions of the observing angle. To model the jet spreading and interaction with the ISM, these results assume the initial jet was a gaussian with $L_{\rm obs}(\theta = 0, t = 11 \ \rm hr) = 8 \times 10^{43} \ \rm erg/s$, and spread out later to a wider, shallower gaussian with the same area and $E_{\rm X}(0.6 \ \rm rad) = 6.5 \times 10^{46} \ \rm erg$, as in Equation~(\ref{eqn:Ex}) and $\Gamma = 6$ \citep{Beniamini_2020}.}
    \label{fig:xray-theta}
\end{figure}

With the initial jet characterized, we then condition the spread jet on the area under the original gaussian and the energy found in Equation~(\ref{eqn:Ex}). The angular dependence of the X-ray energy with the spread jet is 
\begin{eqnarray} \label{eqn:E_x_theta}
    E_{\rm X}(\theta_{\rm v}) &\propto& \left\{ 
    \begin{array}{lll} 
        e^{-\frac{1}{2}\left( \frac{\theta_v}{\theta_{\rm j, spread}/2} \right)^2 } &,& \theta_{\rm v} \leq \theta_{\rm j, spread} \nonumber \\
        \dop{\theta_{\rm v} - \theta_{\rm j, spread}}{3} &,& \theta_{\rm v} > \theta_{\rm j, spread} \nonumber \\
    \end{array} \right. \ \, .
\end{eqnarray}
and for viewing angles beyond the spread jet, the same beaming factor is used but with $\Gamma = 6$ \citep{Beniamini_2020}, as with GRB170817A. This prescription  assumes GRB170817A to follow typical values of GRBs, despite it having a very low peak energy \citep{Goldstein_2017}, thus we are overestimating the on-axis contributions of the afterglow, but our focus is to illustrate the contribution of the afterglow at large angles.

Using the energy described in Equation~(\ref{eqn:E_x_theta}), the lethal distance then depends on the viewing angle in the following way
\begin{eqnarray}
    \label{eqn:xkill_theta}
    \dkill_{\rm X}(\theta_{\rm v}) &=& \left( \frac{E_X}{4\pi \flu^{\text{kill}}_X} \right)^{\frac{1}{2}} \, .\\
    &=& \dkill_{\rm X}(\theta_{\rm v, 0})\pfrac{\dop{\theta_{\rm v} - \theta_{\rm j}}{}}{\dop{\theta_{\rm v,0} - \theta_{\rm j}}{}}^{3/2} \, , \nonumber 
\end{eqnarray}
where $\dkill_{\rm X}$ is as in Equation~(\ref{eqn:Dxkill}) and $\theta_{\rm v, 0} = 0.6 \ \rm rad$. Figure~\ref{fig:xray-theta} shows the angular dependence of the energy and lethal distance while taking into consideration emissions in the temporal range $t_{\rm break, 2}$ to $4\ \rm yr$. As expected, the range at which an afterglow is lethal depends strongly on the direction of the Earth with respect to the jet. In particular, roughly 5 degrees outside the jet the kill distance has already dropped below $1 \ \rm pc$. 

\section{Gamma-Ray Threat} \label{sec:gammaray}

\subsection{$r$-Process Ejecta}

\begin{figure}
    \centering
    \includegraphics[scale = 0.5]{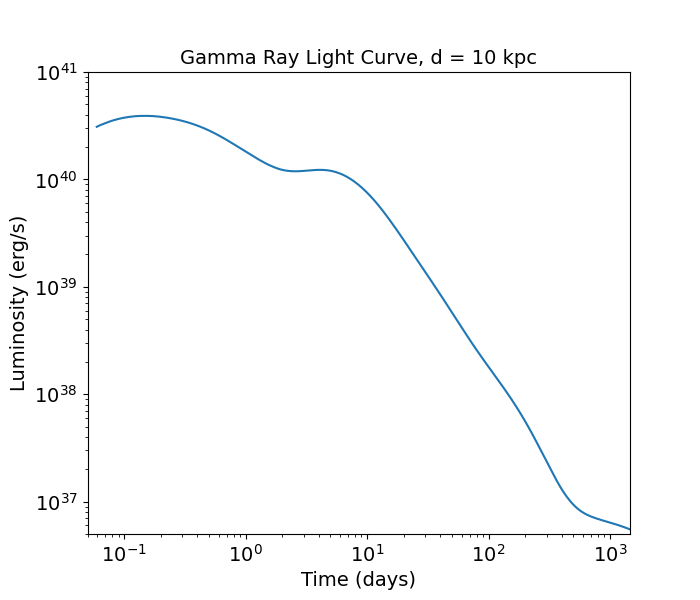}
\caption{The gamma-ray light curve of dynamical ejecta from a BNS merger located at a distance of $10 \ \rm kpc$ with a total {\em r}-process ejecta mass of $0.01 \ \rm M_{\odot}$, during the period before the ozone begins to recover. Gamma-ray photons from both beta decay and the fission of the $r$-process species synthesized in the merger are included. The gamma-ray calculation is adapted from \citet{wang_mev_2020}.}
    \label{fig:gamma-lc}
\end{figure}

BNS mergers are expected to produce $r$-process nuclei copiously, and gamma-ray photons will be emitted from the decays of the exotic, neutron-rich species synthesized in the merger. 
GW170817a was too far away for these gamma rays to be detected, so we rely on theoretical calculations of this emission.
Here we adopt the gamma-ray calculations in \cite{wang_mev_2020} to estimate the gamma-ray threat from the BNS $r$-process ejecta. 
The nucleosynthetic yields of the ejecta are obtained using the nuclear reaction network code PRISM \citep{Mumpower2018} with the nuclear data described in \cite{wang_mev_2020} (the FRDM2012
mass model \citep{FRDM2012}, the FRLDM fission barrier heights \citep{FRLDM},
and the GEF2016 fission yields \citep{GEF2016}, combined with experimental data whenever possible). For individual gamma-ray spectra, we adopt the ENDF/B-VIII.0
database \citep{ENDF8} for experimental beta-decay gamma-ray photons and the 2016 version of the code GEF \citep{GEF2016}, as in \cite{Vassh2019}, for theoretical fission gamma-ray photons. 
The gamma-rays promptly emitted from both beta decays and fission are then propagated through the merger ejecta using the
same radiation transfer methods described in \cite{wang_mev_2020}. 
There are two primary mechanisms by which $r$-process material flows out of a neutron star merger: dynamical ejecta and viscously-driven wind ejecta. For illustration, the resulting gamma-ray light curve for photons in the energy range between $25 \ \rm keV$ and $25 \ \rm MeV$ for dynamical ejecta \citep{Rosswog2013} from a BNS merger at a distance of $10 \ \rm kpc$ that yields $0.01 M_{\odot}$ of {\em r}-process material is shown in Fig.~\ref{fig:gamma-lc}.

As discussed in Section~\ref{sec:xray}, the recovery timescale of the ozone is about 4 years, so we only consider gamma-ray emissions within such a period when assessing the threat of damage. Additionally, we calculate the gamma-ray luminosity starting from $\sim1.5 \ \rm hr$ after the merger, as earlier gamma-ray photons are obstructed by the high opacity of the $r$-process ejecta. 
Thus we estimate the total energy in gamma rays from the dynamical ejecta $E_{\rm ej, \gamma}$ from a merger at distance $\dsn$ to be:
\begin{equation}
    \label{eqn:gamma-e}
    E_{\rm ej,\gamma} = 4\pi D^2 \int \int E_{\gamma}\frac{dN_{\gamma}}{dE_{\gamma}dAdt} dE dt = \int L_{\gamma} dt \, .
\end{equation}
We include emitted gamma-ray photons with energies between $25 \ \rm keV$ and $25 \ \rm MeV$ for luminosity and energy estimates, as orders of magnitude fewer gamma-ray photons are emitted outside this energy range. 
Utilizing the commonly-used critical gamma-ray fluence $\flu^{\rm kill}_{\gamma} = 100 \ \rm kJ \ m^{-2}$ \citep{mel_astrophysical_2011}, the resulting gamma-ray energy and lethal distance are
\begin{eqnarray}
    E_{\rm ej,\gamma} &\approx & 2.1 \times 10^{46} \left( 
    \frac{M_{{\rm ej},r}}{0.01 M_{\odot}} \right)\ \rm erg \, , \label{eqn:ej_energy} \\
    \mathcal{D}_{\gamma, \ \rm ej}^{\rm kill} &\approx & 1.3 \left( 
    \frac{M_{{\rm ej},r}}{0.01 M_{\odot}} \right)^{\frac{1}{2}} \ \rm pc \, . \label{equ:dleth_ej}
\end{eqnarray}
The former is consistent with an estimate based on the total gamma radiation rate from the $r$-process ejecta  $\epsilon_0(\tau)\sim2\times 10^{10} {\rm erg \ \rm g}^{-1} s^{-1} (\tau/\rm day)^{-1/3}$ \citep{Metzger2012,Oleg2020}, which gives $\sim 2.4 \times 10^{46} (M_{{\rm ej},r}/0.01 M_{\odot})\ \rm erg$, well below the total energy released through the $r$ process in mergers \citep[e.g.,][]{Yonglin2021}. The latter is comparable to the X-ray lethal distance calculated in Section~\ref{sec:afterglow}.

Because the gamma-ray photons arise from the decays of the $r$-process radioisotopes synthesized in the ejecta, the total gamma-ray energy or luminosity depends on the overall abundance/mass of the unstable species as well as the composition of the $r$-process radioisotopes. 
The BNS ejecta become optically thin after about 10 days \citep{wang_mev_2020}, when most of the gamma-ray photons have escaped from the ejecta. Therefore, the gamma-ray emission at this later time is expected to scale almost linearly with the mass of the {\em r}-process ejecta: $E_{\rm ej,\gamma} \propto M_{{\rm ej},r}$, which is also indicated in Equation~(\ref{eqn:ej_energy}). The BNS $r$-process ejecta mass normally ranges over $\sim 5\times 10^{-3}-0.1 M_{\odot}$, according to the theoretical estimates in \cite{wang_r-process_2021} and the values suggested by the observations of GW170817 \citep{Cote2018}. 

We note also that the degree of neutron-richness has an impact on the composition of the synthesized {\em r}-process nuclei and the resulting gamma-ray emission \citep{wang_mev_2020}. A smaller initial electron fraction, corresponding to more neutron-rich conditions, leads to the synthesis of heavier $r$-process nuclei like the actinides, which can fission. More unstable nuclei are created in such a robust $r$-process environment, and they decay emitting gamma-ray photons through alpha decay and beta decay as well as fission, resulting in relatively stronger gamma-ray emissions for the same $r$-process ejecta mass \citep[e.g.,][]{Oleg2020, wang_mev_2020}. 
The dynamical ejecta examples illustrated here arise from very neutron-rich conditions with an initial electron fraction of $Y_e \sim 0.015$, which promotes robust fission during the $r$-process. \cite{wang_mev_2020} also calculated the gamma-ray emission for parameterized BNS low-entropy outflow conditions, as could be found in both prompt and wind ejecta \citep{Just2015, Radice2018}, with a range of initial electron fractions $Y_e=0.15$-$0.3$. The gamma-ray luminosity and total energy estimated for these calculations are similar (for $Y_e=0.15$) or much lower (for $Y_e=0.3$) than the dynamical ejecta results, suggesting that the gamma-ray emission could be less threatening for mergers with the bulk of the ejecta having $Y_e > 0.15$, as can occur if the merger remnant is a massive neutron star. 
Indeed, the electron fraction for some component of the BNS ejecta may rise locally to nearly $Y_e = 0.5$ \citep{Shibata_2019} due to irradiation by neutrinos. Thus, there may be gamma-ray emission from iron group radioisotopes like $^{56}$Ni that are produced in $Y_e \sim 0.5$ nuclear statistical equilibrium, though the mass of such ejecta is expected to be much smaller than that of the $r$-process.

\subsection{The GRB Cocoon}

As we are interested in emissions from the BNS merger that are seen off the axis of the sGRB, emission from the GRB cocoon is of interest. Here we consider the prompt sGRB emission scattered by the cocoon to large angles as well as the shock breakout of the cocoon.

\cite{Kisaka_2018} calculated the gamma-ray luminosity for off-axis gamma rays scattered by the cocoon, finding $L_{\rm cocoon, sc} = 10^{47} \ \rm erg / s$ with a duration of $2 \ \rm s$, which is consistent with the observations of GRB 170817A. In the absence of a light curve, we assume the luminosity to be constant for the duration of the scattering emission. This approximation gives us a total energy on the order of $10^{47} \ \rm erg$ and, given that most of the emission is above $10 \ \rm keV$, we use a critical fluence of $\flu^{\rm kill}_{\gamma} = 100 \ \rm kJ \ m^{-2}$ and find a lethal distance of 
\begin{equation}
\label{eq:Dcoc1}
    \mathcal{D}_{\gamma, \rm cocoon, sc}^{\rm kill}(\sim 0.4 \ \rm rad) = 4 \ \rm pc \, . 
\end{equation}

We also considered the emission from the shock breakout of the cocoon. \cite{gottlieb_cocoon_2018} calculated the gamma-ray light curve and spectra for shock breakout in a cocoon of a choked sGRB jet observed at an angle of $\theta = 0.7 \ \rm rad = 40 \ \rm deg$ to obtain a spectrum and light curve consistent with GRB 170817A data. We compute the total energy using a simple geometric integral, which results in a total energy of the order of $10^{47} \ \rm erg$. Given that both the scattered model, discussed above, and the breakout model are means of replicating GRB 170817A, it is expected that the resulting energies are similar. Therefore, we again find the breakout lethal distance to be
\begin{equation}
\label{eq:Dcoc2}
    \mathcal{D}_{\gamma, \rm cocoon, bo}^{\rm kill}(0.7 \ \rm rad) = 4 \ \rm pc \, . 
\end{equation}
However, we caution that there is some uncertainty associated with the breakout spectra used, as there is a range of possible photon energies and cocoon durations \citep{gottlieb_cocoon_2018}.

Despite the brief emission, lasting $< 2 \ \rm s$, the cocoon emission is lethal out to distances greater than that of the afterglow and $r$-process ejecta, for the baseline parameters we have chosen. 

\subsection{Angular Dependence of the Gamma-Ray Emission} \label{sec:gamma-angle}

In the previous Sections, we considered three physically different sources of gamma-ray emission, and now we consider how the observing angle changes the energy received.  Kilonovae are typically assumed to be approximately isotropic, and \cite{Sneppen_2023} found the ejecta from GW170817 to be highly spherical indicating, supporting this approximation. In the case of the $r$-process ejecta, the observing angle has minimal effect on the luminosity measured.   

In contrast to the roughly isotropic {\em r}-process emission, the gamma rays from the coccoon should depend strongly on viewing angle.
\cite{Kisaka_2018} provides an explicit angular dependence for the scattered emission, which we consider here. Additionally, we are more inclined to use the results of \cite{Kisaka_2018} as the viewing angle reported to replicate the observation ($\theta_v \sim 25-30 \deg$) is closer to the typically reported value for GW170817 of $0.4 \ \rm rad \approx 23 \deg$ \citep{Finstad_2018, Abbott_2019}, whereas \cite{gottlieb_cocoon_2018} reported $\theta_v = 0.7 \ \rm  \approx 40 \deg$. However, we acknowledge that the actual emission is likely to be a combination of both scattered and breakout emission.

We use the estimated angular dependence of the energy given in Fig.~3 of \cite{Kisaka_2018}, with a 2-component exponential structure. As done in previous sections, we model the GRB with a gaussian structure and opening angle of $0.1 \ \rm rad$. For a viewing angle $\theta_{\rm v} > \theta_{\rm j}$, the energy is found by
\begin{eqnarray} \label{eqn:Ecocoon_theta}
    E_{\rm cocoon} &\propto& \left \{ 
        \begin{array}{ccc}
           e^{\frac{-\theta_{\rm v}}{\theta_{1}}}  &,& \theta_{\rm v} \leq \theta_{\rm break} \\
           e^{\frac{-\theta_{\rm v}}{\theta_{2}}} &,& \theta_{\rm v} > \theta_{\rm break}
        \end{array} \right. \ \, ,
\end{eqnarray}
where $\theta_{\rm break}$ is roughly $5^{\circ}$ away from the edge of the jet and $\theta_1$ and $\theta_2$ are the characteristic angular sizes of each component. Then, the lethal distance is
\begin{eqnarray} \label{eqn:Dcocoon_theta}
    \dkill_{\rm cocoon}(\theta_v) &=& \left( \frac{E_{\rm cocoon}}{4\pi \flu^{\text{kill}}_{\gamma}} \right)^{\frac{1}{2}} 
     = \dkill_{\rm cocoon}(\theta_{v,0}) \ e^{-(\theta_{\rm v}-\theta_{v,0})/2\theta_i} \, ,
\end{eqnarray}
where $\dkill_{\rm cocoon}$ is as in Equation~(\ref{eq:Dcoc1}) and (\ref{eq:Dcoc2}) with  $\theta_{v,0}=0.4 \ \rm rad$ and $\theta_i = \theta_1$ or $\theta_2$ depending on the viewing angle, as seen in Equation~(\ref{eqn:Ecocoon_theta}).

\begin{figure}
    \centering
    \includegraphics[width=\textwidth]{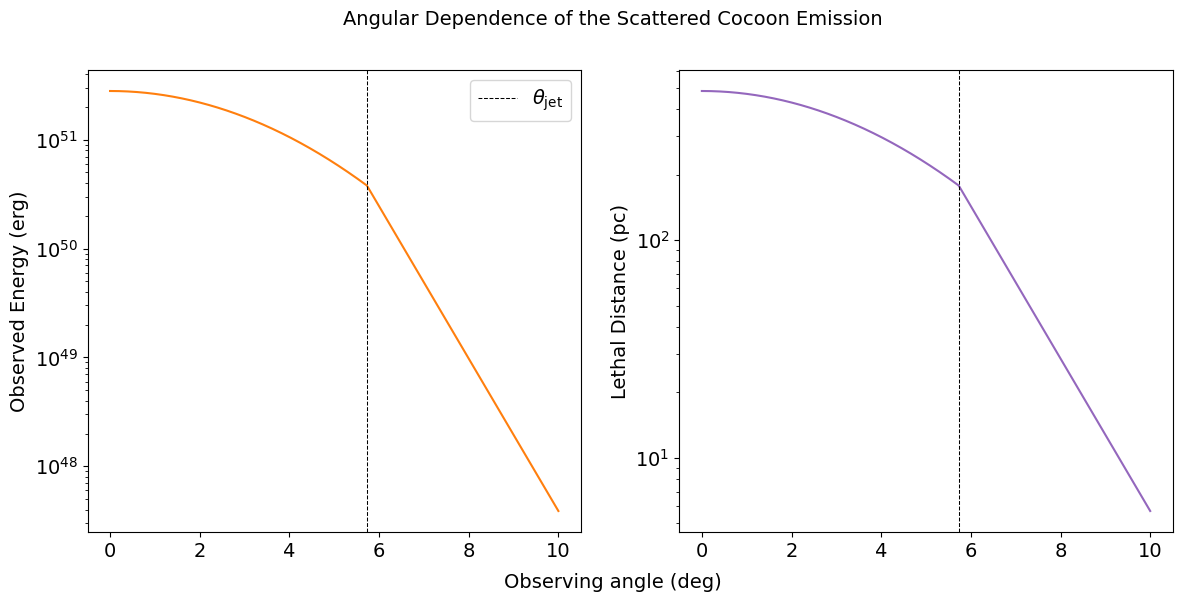}
    \caption{The left and right panels show the energy from scattered cocoon gamma rays and the associated lethal distance, respectively, as functions of the observing angle. These results assume a gaussian jet structure and $E_{\rm obs}(\theta = 0.4 \ \rm rad) = 10^{47} \ \rm erg$.} 
    \label{fig:cocoon-theta}
\end{figure}

The resulting angular dependence for the energy and lethal distance is shown in Fig.~\ref{fig:cocoon-theta}. 
Despite the significant decrease in the cocoon energy, it is several orders of magnitude greater than the energy emitted by the sGRB seen in Fig.~\ref{fig:GRB-theta} for a viewing angle of $10^{\circ}$.  We also see that the strong dropoff with viewing angle means that outside of $\sim 10^\circ$ the kill distance from the coccoon falls below that of the {\em r}-process ejecta. Thus we see that at small viewing angles, the coccoon emission dominates, while at large angles the {\em r}-process gamma rays are the main threat.

The photon damage from a kilonova starts almost immediately with the arrival of its first light, and the effects will persist on a $\sim 4$ year timescale until the stratospheric ozone can be replenished.  The biosphere will then have respite for hundreds to thousands of years, but then a new hazard will emerge, with the arrival of cosmic rays.

\section{Cosmic-Ray Threat} \label{sec:cosmicray}

Over years to millennia after the BNS merger, the ejecta create an interstellar bubble.  The BNS merger explosion releases up to $10^{51}$ erg of energy, which launches a strong shock that sweeps up interstellar matter. The resulting structure--a BNS remnant--should be closely analogous to the SN remnants that dot the Galactic plane \citep{Green2019}, and are born with a similar energy releases \citep[e.g.,][]{Montes2016, Wu2019}.  Supernova remnants are engines of cosmic-ray acceleration \citep[e.g.,][]{Ackermann_2013}, and thus we expect that merger events also create cosmic rays, which are potentially harmful to nearby biospheres.

Much like the photon emission described in Section \ref{sec:ionizing-photons}, cosmic rays also deplete the ozone layer through the creation of nitrate ions \citep{Gehrels_2003, Atri_2014, Melott_2017}. Additionally, ionization of the atmosphere can lead to increases in lightning strikes and therefore wildfires \citep{Atri_2014}. Cosmic rays also produce secondaries when they interaction with the atmosphere \citep{Ferrari_2009}. Among these are high-energy muons, which can penetrate the Earth's surface and several hundred of meters deep into water and are damaging to biota \citep{Atri_2014, Dar_1998, Juckett_2009}. As a result, even organisms in caves or in the ocean depths can be irradiated by these harmful particles \citep{Ferrari_2009, Atri_2014}. Significant cosmic-ray irradiation from a kilonova could therefore have drastic consequences for life on Earth.

Work to date on cosmic-ray damage has focused on nearby SNe; here we are concerned with KNe and so need to explore how BNS production of cosmic rays resembles or differs from the SN case.
While a KN is may well be more aspherical than many SN explosions,  the BNS remnant should eventually lose memory of its initial geometry and become more spherical and possess similar properties to SN remnants. \cite{Montes2016} found that the size of the BNS remnant is similar to that of a SN remnant at the time of shell formation, or the end of the Sedov-Taylor phase.

In the context of SNe, the damage is most significant at this phase because, as the forward shock passes by, the cosmic rays that were trapped inside the remnant and possess $\sim 10 \%$ of the blast energy can now reach Earth \citep{mel_astrophysical_2011}. 
Given the late-time similarities to SN remnants, we would expect the kill distance to be similar to that for a SN, if the blast energy going into cosmic rays is similar that of a canonical SN.

Within a model of cosmic-ray acceleration we can estimate the cosmic-ray kill distance and its dependence on the blast energy.
The time and spatial history of cosmic-ray acceleration in a remnant is a subject of active research, as is the nature of cosmic ray escape into the interstellar medium, and thus a complete picture remains uncertain.  But in diffusive shock acceleration, charged particles are accelerated in repeated crossings of the magnetized shock, and then advected downstream, into the remnant \citep[e.g.][]{Schure2012}.  
Recent work suggests that the remnant initially confines the newborn cosmic rays, until over time they diffusively escape.  Higher-energy particles leave first, while their lower energy counterparts--which make up the bulk of the particles and of the energy--leave last \citep[e.g.][]{TDP2012,Bell2013,Zweibel2013,Celli2019,Brose2020}.

 Thus, in this picture, the cosmic-ray damage for BNS mergers (and for SNe) only begins around the time the blast wave arrives at Earth.  The blast arrival timescale depends on the ambient density and the explosion energy, but will occur hundreds to thousands of years after the explosion.  Thus the cosmic-ray damage will begin substantially later than the initial photon damage we have studied in earlier sections.  Moreover, the cosmic rays will linger until they finally escape the remnant, so the damage to the atmosphere and biosphere will be sustained, also for thousands of years.  

Within this scenario, the cosmic-ray density builds up to maximum and then declines, over periods of thousands of years. We model this in a simple way as follows.
For an explosion with blast energy $E_{\rm blast}$,
let some fraction of this energy go into cosmic rays.  We further imagine that a portion of the cosmic rays remain inside the remnant; we let
$\eta_{\rm cr}$  be the fraction of blast energy that goes to these confined cosmic rays,
so that they have a total energy $E_{\rm cr} = \eta_{\rm cr} E_{\rm blast}$.
The average confined cosmic-ray energy density is thus
$u_{\rm cr} = 3E/4\pi r^3$,
where $r$ is the remnant radius.

\citet{Gehrels_2003} argue that cosmic rays will cause dangerous ozone loss if their energy flux 
reaches $F_{\rm crit} \sim 5 \ \rm erg \ cm^{-2} \ s^{-1} = 150 \ kJ \ m^{-2} \ yr^{-1}$. This is about 100 times the present-day flux.  
For isotropic relativistic cosmic rays,
their energy flux $F_{\rm cr}$ and energy density $u_{\rm cr}$ are related by $F_{\rm cr} = u_{\rm cr} c/4$.  The confined cosmic-ray flux matches the critical level $F_{\rm cr} = F_{\rm crit}$
at the radius
\begin{eqnarray}
\dkill_{\rm cr} & \simeq & \pfrac{3 \eta_{\rm cr} E_{\rm blast} c}{16 \pi F_{\rm crit}}^{1/3} \\
\label{eq:DkillCR}
& = & 11 \ {\rm pc} 
 \ \pfrac{5 \ \rm erg \ cm^{-2} \ s^{-1}}{{F}_{\rm crit}}^{1/3}
 \ \pfrac{\eta_{\rm CR}}{0.1}^{1/3}
 \ \pfrac{E_{\rm blast}}{1 \ \rm foe}^{1/3} \, ,
\end{eqnarray}
where $1 \ \rm foe = 1 \ \rm Bethe = 10^{51} \ \rm erg$
is the canonical supernova blast energy.
We note that this estimate is comparable to that found in \citet{Gehrels_2003}.   We note also that the blast energy dependence is rather weak,
with $D_{\rm kill,cr} \propto E_{\rm blast}^{1/3}$.

Equation (\ref{eq:DkillCR}) shows that the cosmic-ray kill distance depends on the total blast energy.  For BNS mergers, this is less well known than for SNe, with estimates for GW170817 in the range 0.1-1 foe \citep[e.g.]{Cowperthwaite_2017, Kilpatrick_2017, troja_thousand_2020} and BNS mergers, generally, fall in the range of 0.3-2 foe, depending on the mass of the constituent neutron stars \citep{metzger_kilonovae_2020}.  

In closing this Section, we return to the question whether BNS mergers indeed create blast waves similar to SNe at late times.
In the case of GW170817, the ejecta possessed a speed of $v = 0.1 - 0.3 \ \rm c$ \citep[e.g.]{Drout_2017, Cowperthwaite_2017, Villar_2017, Waxman2018, Bulla_2019, Shibata_2019}. Given this speed and the known distance of $40 \ \rm Mpc$ to the merger, it is possible to estimate the remnant's angular size on the sky. The angular diameter for an object of diameter $d$ at a distance of $D$ is
\begin{eqnarray}
    \delta &=& \frac{d}{D} = \frac{2vt}{D} \nonumber \\
    & = & 0.32 \ {\rm mas} \ \pfrac{t}{1 \, \rm yr} \pfrac{v}{0.1 \, c} \pfrac{40 \, \rm Mpc}{D} \, 
\end{eqnarray}
where $1  \ \rm mas = 10^{-3} \ arcsec$, and where we assume the blast undergoes free expansion at speed $v$.
It has been nearly six years since the merger, and for the slower ejecta with $d = vt$, the angular diameter is approximately
\begin{eqnarray}
    \delta_{\rm now} &=& 1.9 \ \rm mas \, ,
\end{eqnarray}
and in ten years, 
\begin{eqnarray}
    \delta_{10} &=& 3.2 \ \rm mas \, .
\end{eqnarray}
This size is similar in scale to the supernova remnants observed by \cite{Varenius_2019} with very long baseline interferometry. If the merger remnant is resolved, then by conservation of surface brightness, since supernova remnants have been observed and merger remnants are very similar, then the merger remnant should also be bright enough to observe.

\section{Merger Rates and Galactic Threat Assessment} \label{sec:rates}

With the threat from each emission source quantified, we will now summarize the previous results and combine them to determine a single value for the lethal distance of BNS mergers. This will then be used to assess the threat BNS mergers could have on life as well as a comparison with other dangerous astrophysical phenomena.

\subsection{Combined Damage and Overall Kill Distance}

In the previous sections, we have characterized the threat imposed by the various components of emissions from BNS mergers. Table~\ref{tab:summary} lists the energy and associated lethal distance for each of the emission components considered in Sections \ref{sec:xray}, \ref{sec:gammaray}, and \ref{sec:cosmicray}. 

Given that the X-ray and gamma-ray emissions considered thus far all occur while the impulse approximation is still valid, all of these emission components are depleting the ozone simultaneously. For a GW170817-like viewing angle $\theta_v \approx 0.4 \ \rm rad$ \citep{Finstad_2018, Abbott_2019, metzger_kilonovae_2020}, the lethal distance induced by both X-ray and gamma-ray photons can be found using Equation~(\ref{equ:dleth-tot}):
\begin{eqnarray}
    \dkill_{\rm net,photon}(\theta_{\rm v}) &=& \sqrt{ \sum_{\rm b} {\dkill_{\rm b}}^2 } \\
    &=& \sqrt{ (\dkill_{\rm X}\dop{\theta_{\rm v} - \theta_{\rm j}}{2})^2 + {\dkill_{\rm \gamma, ej}}^2 + (\dkill_{\rm cocoon}e^{-\theta_{\rm v}/\theta_i})^2 } \nonumber \\
    &=& 6.6 \ \rm pc \, , \nonumber 
\end{eqnarray}
where the final numerical value combines the results from the individual components using Equation~(\ref{equ:dleth-tot}) with the appropriate weights, evaluated at the benchmark parameters discussed in each section above.
For these parameters, the photon emission impacts a smaller range than the cosmic rays that arrive at later times. Note that the lethality of the sGRB beam is not included in Table~\ref{tab:summary}: our focus here has been the off-axis emission.

\begin{table}[htb] 
    \centering
    \caption{   
    Baseline Values for Emission Energies and Lethal Distances
    \label{tab:summary}
        }
    \begin{tabular}{c|c|c|c}
 Emission Component & Band & $E_{\rm b} \ \rm (10^{47} \ erg)$ &  $\mathcal{D}_{\rm b}^{\rm kill} \ (\rm pc )$ \\
\hline
GRB Afterglow${}^*$ & X-ray  & 11 & 4.8 \\
$r$-Process Ejecta & $\gamma$-ray & 0.21 & 1.3 \\
Cocoon${}^*$ & $\gamma$-ray & 2 & 4 \\
\hline
Photons${}^*$ & X $+$ $\gamma$ & 5.2 & 6.6 \\
\hline \hline
Remnant & Cosmic ray & & 11 \\
\hline 
    \end{tabular} \\
    \tablenotetext{*}{The afterglow and cocoon emissions depend on the viewing angle; the results are shown for $\theta_{\rm v} = 0.4 \ \rm rad$ for GW170817 \citep{Finstad_2018, Abbott_2019, metzger_kilonovae_2020}, and the parameter scalings are included in Equation~(\ref{eqn:xkill_theta}) and (\ref{eqn:Dcocoon_theta}).}
\end{table}

Comparing the kill distances found in Equations~(\ref{equ:dleth_ej}), (\ref{eq:Dcoc1}), (\ref{eq:Dcoc2}) and (\ref{eq:DkillCR}), and shown in Table~\ref{tab:summary} with the baseline parameter choices, the cosmic-ray kill distance is the largest, and indeed the associated damage may be the most severe due to the long exposure of the cosmic rays.  But we see that the different estimates are comparable.
Hence there is not a clearly dominant kill mechanism, especially when accounting for the uncertainties in each of our estimates. As we found in Sections~\ref{sec:xray-angle} and \ref{sec:gamma-angle}, the energy emitted by the afterglow and cocoon, respectively, are the most sensitive to the observing angle, and the numbers shown in Table~\ref{tab:summary} are conservative, in the sense that their $\mathcal{D}_{\rm b}^{\rm kill}$ would be sharply reduced at larger angles from the GRB. Both components are also influenced by the energy of the jet, and typical values span five orders of magnitude \citep{Salafia_2020}. This results in the lethal distances being uncertain within an of order of magnitude. Additionally, the cocoon emission is impacted by numerous other factors such as the cocoon Lorentz factor and mass as well as additional jet parameters including the jet Lorentz factor and opening angle \citep{Kisaka_2018}, so this is an underestimate of the uncertainty. The $r$-process ejecta and cosmic rays are isotropic features and will be dominate at large viewing angles, but also have intrinsic variations. The $r$-process emission is proportional to the mass ejected, which can range up to an order of magnitude higher or lower than assumed here. This range results in the energy being uncertain up to an order of magnitude and the lethal distance by a factor of $\sim 3$. Similarly, the cosmic rays depend on the merger blast energy which can vary up to an order of magnitude resulting in a factor of $\sim3$ uncertainty in the lethal distance. Overall, it is possible to envision cases in which any of the emission components are the most threatening.

Moreover, if the merger is close enough, then like a supernova it can bring significant damage in multiple waves.  First the ionizing photons will arrive, but will be gone within a year or so.  Survivors of this insult and their descendants will have a respite for thousands of years, but then the cosmic rays will arrive, initiating a second and sustained wave of damage.  The species best suited to survive would be those less susceptible to the radiation effects, perhaps by having a hardy constitution or by living deep in shielded environments.

\subsection{Rates of Nearby Mergers and Threat Assessment}

In order to estimate the rate of lethal binary neutron star mergers, we assume that the distribution of galactic merger progenitors can be described by a double exponential disk model with scale radius and height. Ranges of values for the scale height ($500-1400 \ \rm pc$) and radius ($2-5 \ \rm kpc$) \citep{Gu_2019} have been given in the literature: see, e.g., \citet{Yaz_2010, Chang_2011, Jia_2014, Wan_2017}. The rate is sensitive to the selection of the scale height and radius, but the specific values do not change the overall conclusions we draw below. We use  $h_0 = 800 \ \rm pc$ and $R_0 = 2.4 \ \rm kpc$, respectively, which correspond approximately to the distribution of long-lived stellar systems within the Milky Way Galaxy \citep{Adams_2013, Girardi_2005, Murphey_2021}. 

The average rate of lethal mergers within a radius $r_{\rm BNS}$ can then be approximated by:
\begin{eqnarray}
    \label{equ:rate}
    \Gamma (r) &\approx & \frac{4\pi r^3}{3} \ \dot\mathcal{N}_{\rm BNS} \ \rho_{\rm BNS}(R_{\odot}, h_{\odot}) \\ 
    &= & \frac{4\pi r^3}{3} \dot\mathcal{N}_{\rm CC} \left( \frac{\mathcal{R}_{\rm BNS}}{\mathcal{R}_{\rm CC}} \right) \frac{e^{-R_{\odot}/R_{0}}e^{-h_{\odot}/h_{0}}}{  4\pi R_{0}^2h_{0}} \, , \nonumber 
\end{eqnarray}
where the probability density of Galactic mergers $\rho_{\rm BNS}(R_{\odot}, h_{\odot})$ is normalized such that $\int \rho_{\rm BNS} \ dV =1$ when evaluated at the observer's position, which is taken to the the location of the Sun in galactocentric coordinates.  The galactic rate of BNS mergers $\dot\mathcal{N}_{\rm BNS}$ can be determined by comparing to the rate of core collapse supernovae (CCSN) $\dot\mathcal{N}_{\rm CC}$. CCSN are a well-understood population, and we assume that the ratio of volumetric rates of BNS, $\mathcal{R}_{\rm BNS}$, and CCSN, $\mathcal{R}_{\rm CC}$, is the same as the ratio of the galactic rates. This results in $\dot\mathcal{N}_{\rm BNS}  = \dot\mathcal{N}_{\rm CC} \ ( \mathcal{R}_{\rm BNS} / \mathcal{R}_{\rm CC} )$.

Following the prescription outlined in \cite{brunton_x-ray_2022}, we adopt the values of $R_{\odot} = 8.7 \ \rm kpc$, $h_{\odot} = 20 \ \rm pc$, and $\dot{\mathcal{N}}_{\rm CC} = 3.2^{+7.3}_{-2.6} \ \rm events/century$, respectively. In addition, the following values are adopted: $\mathcal{R}_{\rm CC} = (1.25 \pm 0.50) \times 10^{-4} \ \rm Mpc^{-3} \ yr^{-1}$ (\citealt{Lien2010}, consistent with \citealt{Taylor2014,Cappellaro2015}), and $\mathcal{R}_{\rm BNS} = 200^{+309}_{-148} \ \rm Gpc^{-3} \ yr^{-1} = 2^{+3.09}_{-1.48} \times 10^{-7} \ \rm Mpc^{-3} \ yr^{-1}$ \citep{nitz_2023}. Thus the ratio of the rates per unit volume of the local universe of BNS mergers to CC events, which we take to be also the ratio of the rate within in the Galaxy of BNS mergers to CC, is $1.8^{+2.8}_{-1.3} \times 10^{-3}$. 

Evaluating Equation~(\ref{equ:rate}) at the greatest lethal distance found, which was associated with the cosmic ray threat, $r = r_{\rm BNS} = 10 \ \rm pc$, the rate of lethal events is 
\begin{equation}
\Gamma_{\rm BNS}(r_{\rm BNS}) = 10^{-3.9^{+1.2}_{-0.5}} \ \rm events/Gyr \left( \frac{r_{\rm BNS}}{11 \ \rm pc}\right)^3  \, ,
\end{equation}
and the mean recurrence time is $\Gamma_{\rm BNS}(r_{\rm BNS})^{-1} = 10^{3.9^{+0.5}_{-1.2}} \ \rm Gyr \approx 500 t_{0}$, where $t_{0} = 13.8 \ \rm Gyr$ is the age of the Universe \citep{Planck2020}.  With such a long recurrence time we see that KNe do not pose a likely threat for the Earth.

The KN threat is substantially small than that of SNe, despite the two having similar kill distances.  The reason for this difference arises in part due to the ratio of event rates. Galactic binary neutron star mergers are much rarer than CCSNe, and they are much less of a threat when viewed off-axis. 
In addition, we expect the BNS population to be old and thus to have a large scale height, diluting their density near the disk midplane.
Given the intrinsic rarity and dilute nature of KNe, the only way for these events to be more threatening than the CCSNe would be to have a much larger lethal distance that would impact a greater volume, which, as we have determined above, is not the case. 

Moreover, there are other astrophysical phenomena that are more threatening to life on Earth, such as CCSNe and on-axis GRBs as mentioned in Section \ref{sec:intro}. By way of comparison, CCSNe, long GRBs, and sGRBs are estimated to have lethal recurrence times of $\sim 1$ \citep{brunton_x-ray_2022}, $\sim 1$ and $\sim 0.3 \ \rm Gyr$ \citep{mel_astrophysical_2011}, respectively, all of which are several orders of magnitude more likely to occur than a lethal BNS merger. An AGN, on the other hand, is not a real threat to life on Earth. The Earth's position is sufficiently far from the Galactic center, even if the Milky Way were currently harboring an AGN \citep{Gonzalez2005}. These events are not threats of major concern on a human timescale. 

Moreover, we recall that solar flares, while influencing a much smaller range than the above-mentioned phenomena, have been known to disturb the atmosphere and interfere with modern-day technological infrastructure, which could threaten human lives. A recent such event was a power outage in Quebec in 1989 \citep{Allen1989}, and the Carrington Event was a previous powerful solar event that led to fires along telephone wires \citep{Carrington1859, Loomis1861, lakhina2004, Noy2022}. \cite{Riley2017} estimate that a similar event could occur within the next $100$ years, which could be devastating for current technology. It has also been estimated that the Sun has the ability to produce ``superflares" that can induce mass-extinction events on Earth with a recurrence time of $20 \ \rm Myr$ \citep{Lingam_2017}. Similarly, impactors are threatening events that could happen on human timescales.  \cite{Stokes2003} finds every $1500$ years on average an impactor event will fatally injure at least 1 person. Impactors with diameters above $1 \ \rm km$ are expected to occur once every $5000$ years and be capable of destroying areas as large as states or countries, and NASA has developed a program for planetary defense to counter such a threat \citep{Morrison_1992}. On a longer timescale, it has been established that the impactor that created the Chicxulub crater caused a mass extinction \citep{Schulte_2010}.

While a BNS merger seems unlikely to cause a mass extinction, it is of interest to consider how it could disrupt a technological civilization.  It is possible that the gamma-ray emission from the cocoon shock breakout or scattering could induce an electromagnetic pulse (EMP) in the atmosphere leading to damage to electronic equipment. The short but intense gamma-ray emission would ionize atmosphere in the hemisphere facing the event, and the free electrons would begin to follow the Earth's magnetic field lines \citep{Glasstone_1997}.  The separation between the electrons and the corresponding ions produces a strong electric field that results in currents in electrical systems that are much greater than they are built to withstand. This could potentially lead to fires in electrical wires, like the Carrington Event and other major solar flares. However, this danger could only occur if the BNS were sufficiently close, and it is likely that the impact from this would be negligible compared to the future incoming cosmic rays. Overall, BNS mergers pale in comparison to other harmful astrophysical events.

It is interesting to note that if a BNS merger were to occur at $\sim10 \ \rm pc$, its apparent magnitude would be equal to its absolute magnitude. With this in mind, one may consider what a nearby BNS merger might look like to an observer on Earth.  Here we consider the V-band magnitude as a proxy for the naked eye, as in \cite{Murphey_2021}. Although the optical kilonova emission has no impact on the ozone layer, a dangerous kilonova would present a spectacular display for the naked-eye. GW170817 had a peak absolute magnitude of $M_V \approx -15.5 \ \rm mag$ \citep{Drout_2017, Villar_2017} and, following the ``best fit" model from \cite{Villar_2017}, after a month had an absolute magnitude $M_V \approx -7 \ \rm mag$. This means that for over a month the kilonova was brighter than the full moon, which has an apparent magnitude of $m_{V, \leftmoon} = -12.71 \pm 0.06$ \citep{Martynov_1959} and is visible during the day - a typical limiting magnitude is $m_{\rm lim, day} \approx -3$ \citep{Weaver_1947} - for an entire month. For comparison, the afterglow from the GRB does emit in the optical, the peak magnitude of GRB 170817A's afterglow was $m_{\rm AB, afterglow} = 26.6$ \citep{Fong_2019} when observed with HST F606W. Therefore, at a distance of 40 Mpc, its absolute peak magnitude was near $-6.4$,  so it would be comparable to the kilonova.

Thus far, we have considered only BNS mergers; we now turn to binary black holes and black hole-neutron star (BHNS) mergers.  For binary black holes, it is expected that little or no matter is involved, and hence little electromagnetic (EM) emission. 
Thus we expect BBH mergers are unlikely to induce extinctions anywhere \citep{metzger_kilonovae_2020}. In the cases of BHNS mergers, there is a range of possible ejecta masses depending on the parameters of the two compact objects \citep{Shibata2016, Shibata_2019, metzger_kilonovae_2020}. In cases where the neutron star is sufficiently compact and massive, the neutron star's self-gravity will keep it intact as it falls into the black hole without any emission  \citep{Shibata2016}. For the merger to produce an EM counterpart, the black hole needs to tidally disrupt the neutron star \citep{Shibata2016, Barbieri_2020, Fragione_2021}. The largest possible ejecta mass for BHNS mergers is estimated to be ${\cal O}(0.1) \ \rm M_{\odot}$, which would correspond to strongly luminous emission, and is greater than the ${\cal O}(0.01) \ \rm M_{\odot}$ expected for BNS mergers. However, for this to happen, the black hole must be spinning rapidly and/or the neutron star equation of state should be very stiff \citep{Kyutoku_2015, Fragione_2021}. These conditions for significant emission are quite restrictive and, based on current studies, both of these conditions are unlikely \citep{Abb_2021, Abb_2018}. Even if every BHNS merger had an EM counterpart, the volumetric rate of these mergers is $\mathcal{R}_{\rm BHNS} = 19^{+30}_{-14} \ \rm Gpc^{-3} \ yr^{-1}$ \citep{nitz_2023}, which is an order of magnitude smaller than $\mathcal{R}_{\rm BNS}$. This rate combined with the expected low luminosity compared to BNS mergers means that BHNS mergers represent a minimal threat.

We conclude that mergers viewed off-axis are unlikely to ever impact life on Earth in any significant way, and they are also unlikely to disrupt potential life in other parts of the Milky Way. \cite{Lineweaver_2004} describes four components of the Galaxy, and the two that are most likely to contain BNS merger progenitors, the halo and thick disk, are too metal-poor to harbor planets with life. We note also that if there were to be life in the Galactic bulge, which harbors the highest concentration of stars, supernovae are the main threat. 

\section{Discussion and Conclusions} \label{sec:discussion}

We have investigated several off-axis emission components of BNS mergers, which included the GRB afterglow and cocoon, as well as dynamical ejecta and the late-time BNS remnant. We found that cosmic rays are most threatening emissions and are potentially lethal out to $\sim 10 \ \rm pc$, similar to the typical value for a core-collapse supernova, 
due to the similar natures of the large-scale remnants. The rarity of binary neutron star mergers combined with a small range of lethality means that they are probably not important threats to life on Earth. We find that the mean recurrence time of lethal mergers at the location of the Sun is much larger than the age of the Universe. However, even if it never induced a mass extinction, a nearby kilonova event would be visible on Earth. It would likely disrupt technology soon after the merger and remain bright in the sky for over a month.

In Section \ref{sec:rates}, we considered various astrophysical threats to biospheres and of these, we believe BNS mergers are the least threatening. Based on the frequency and potential damage done, the threats in order of most to least harmful are: solar flares, impactors, supernovae, on-axis GRBs, and lastly off-axis BNS mergers. Even though it is unlikely life would be extinguished by a merger, a nearby event could be fatal to some. If sufficiently close, the brief cocoon emission would ionize the atmosphere causing a hemisphere-wide electromagnetic pulse. Additionally, astronauts on the ISS or the Moon would be irradiated by the gamma rays and cosmic rays with no protection from the atmosphere. The cosmic rays would cause astronauts to experience flashes of light while their eyes are closed, as was experienced by Apollo astronauts \citep{Osborne_1975}. For those on Earth, the muons produced in the atmosphere would be difficult to avoid and they have been found to cause mutations and birth defects \citep{Dar_1998, Juckett_2009, Melott_2017}.

It is important to note that all of the aforementioned analysis is based on a combination of theory and the single event GW170817/GRB 170817A. 
Further multi-messenger observations of BNS mergers will greatly improve the study conducted here. {Moreover, we anticipate other developments that will enable more accurate modeling of BNS mergers and the composition of their ejecta, including advances in astrophysical simulations (e.g., \cite{Zappa+2023,Foucart+2023,Radice:2023zlw,Curtis:2022hjy}), nuclear experiments that will probe the physics of dense matter (\cite{Sorensen+2023}) and the properties of increasingly exotic nuclei (e.g., \cite{Crawford+2022,Orford+2022}), and improvements to the theoretical treatment of microphysics such as neutrino emission, absorption, and oscillations (e.g., \cite{Gizzi:2021ssk,Grohs+2023,Balantekin+2023}).} Current and future surveys such as LSST, Roman, ZTF, and ATLAS will enable greater studies of these rare phenomena. Estimates by \cite{Scolnic_2017} indicate that these surveys should see $\sim69, 16, 11$, and 8 kilonovae, respectively. {In addition, future gamma-ray observations of kilonovae by next-generation MeV gamma-ray observatories such as COSI \footnote{\url{https:// cosi.ssl.berkeley.edu}} and MeVGRO \footnote{\url{https://indico.icranet.org/event/1/contributions/777/}} will provide more direct information about the gamma-ray fluence.}
These experimental, theoretical and observational advances will enable the possible threat from BNS mergers to be constrained more accurately.

\section*{Acknowledgments}

We would like to acknowledge the insightful conversions with Gautham Narayan that improved much of the work done here. The work of J.E. was supported partly by the United Kingdom STFC Grant ST/T000759/1.
The work of B.D.F.~was supported by
 the U.S. National Science Foundation (NSF) under grant number AST-2108589. The work of R.S, G.C.M and Z. L. was supported under contract LA22-ML-DE-FOA-2440.
The work of R.S. and G.C.M. was supported by the NSF under grant number PHY-2020275 for the Network for Neutrinos, Nuclear Astrophysics, and Symmetries (N3AS) and the U.S. Department of Energy (DOE) under contract DE-SC00268442 (ENAF). This work was supported as well by the U.S. DOE under grant numbers DE-FG02-95-ER40934 (RS) and  DE-FG02-02ER41216 (GCM).
The work of X.W. was
supported by National Key R\&D Program of China
(2021YFA0718500) and the Chinese Academy of Sciences
(Grant No.\ E329A6M1). The authors certify that AI was not used in writing this paper.

\bibliographystyle{aasjournal}
\bibliography{kne_srcs}{}

\end{document}